\documentclass[mathleft
]{an}
\usepackage{graphicx}
\usepackage{times}
\begin{document}

\Pagespan{001}{003}
\Yearpublication{2009}%
\Yearsubmission{2008}%
\Month{00}%
\Volume{000}%
\Issue{00}%

\title{High Frequency Peakers: The Faint Sample}

\author{C. Stanghellini\inst{1}\fnmsep\thanks{Corresponding author:
  \email{cstan@ira.inaf.it}\newline}
\and  D. Dallacasa\inst{1,2}
\and  M. Orienti\inst{1,3}
}
\titlerunning{High Frequency Peakers: The Faint Sample}
\authorrunning{Stanghellini et al.}
\institute{
Istituto di Radioastronomia -- INAF, Bologna, Italy
\and 
Dipartimento di Astronomia - Universit\`a di Bologna, Italy
\and 
Instituto de Astrofisica de Canarias, Tenerife, Spain}

\received{08 Dec 2008}
\accepted{18 Dec 2008}
\publonline{later}


\abstract{%
We present a sample of sources with convex radio spectra peaking at frequencies above a few GHz, known as "High Frequency Peakers" (HFPs). A "bright" sample with a flux density limit of 300 mJy at 5 GHz has been presented by Dallacasa et al. (2000). Here we present the "faint" sample with flux density between 50 and 300 mJy at 5GHz, restricted to the area around the North Galactic Cap, where the FIRST catalogue is available. The candidates have been observed with the VLA at several frequencies ranging from 1.4 to 22 GHz, in order to derive a simultaneous radio spectrum. The final list of confirmed HFP sources consists of 61 objects.}

\maketitle

\section{Introduction}

In the framework of powerful radio-galaxy evolution,
GPS and then CSS radio sources are nowadays considered the early
stages, as the radio emitting region grows and expands within the
interstellar matter of host galaxy, before plunging into the
intergalactic medium to originate the extended radio source population
(Orienti and Dallacasa 2008,2009).

Most of the samples of powerful CSS and GPS radio sources consist of
sources with turnover frequencies ranging from about 100 MHz to about
5 GHz.

Objects with turnover frequencies above 5 GHz would
represent {\it smaller} and therefore {\it younger} radio sources. We call
these sources ``High Frequency Peakers'' (HFPs).

\section{Candidate High Frequency Peakers}
\label{structure}

From the 
cross correlation of the 87GB catalogue at 4.9 GHz and the NVSS
catalogue at 1.4 GHz we selected the sources with inverted spectra
with a slope steeper than 0.5
($S\propto\nu^{-\alpha}$).
We defined {\bf two} samples of candidates based on the flux density in the 87GB 
catalogue: the ``{\bf bright}''
sample (S$_{5GHz}>300$mJy) covering $0^\circ <$ DEC $<+75^\circ$ but excluding objects
projected on the galactic plane ($\mid{b_{II}\mid}>10^\circ$) and the ``faint'' sample
(50mJy $<$ S$_{5GHz} <300 $mJy) restricted to an area around the northern galactic cap
($07^h:20^m < $RA$< 17^h:10^m$, $22.5^\circ <$DEC$<  57.5^\circ$) covered by the FIRST
survey as well (Becker et al. 1995).

Here we present the ``faint'' sample. A description of the ``bright''
sample and the methods used in the selection can be found in Dallacasa et al. (2000).
\begin{table}[!h]
\caption{VLA observations and Configurations. The total observing time
(column 3) is inclusive of the scans on the ``bright'' HFP candidates. }
\begin{center}
\begin{tabular}{cccc}
\hline
\hline
Date        & Conf. & Obs. & code \\
            &       & Time &\\
\hline
07~Nov~1998 &  BnC  & 150  &  a\\
14~Nov~1998 &  BnC  & 150  &  b\\
19~Dec~1998 &   C   & 240  &  c\\
14~Jun~1999 &  AnD  & 420  &  d\\
21~Jun~1999 &  AnD  & 180  &  e\\
25~Jun~1999 &   A   & 120  &  f\\
25~Sep~1999 &   A   & 240  &  g\\
15~Oct~1999 &  BnA  & 240  &  h\\
25~Feb~2000 &  BnC  & 240  &  i\\
\hline
\end{tabular}
\end{center}
\end{table}

\begin{table*}[!t]
\caption{Faint sample. Optical info is based on the NED or is derived from the SDSS;  redshifts in parantheses are photometric. For the coordinates only seconds of time (RA) and arcseconds (DEC) are given }

\begin{center}
\begin{tabular}{lcccrrrrrrrrrr}
\hline
$name$&$ID$&z&ra-dec&
$S_{1.4}$&$S_{1.7}$&$S_{4.5}$&$S_{5.0}$&$S_{8.1}$&$S_{8.5}$&$S_{15}$&$S_{22}$&$\nu_m$&$S_m$\\
         &       & &$_{J2000}$&
\multicolumn{8}{c}{$_{mJy}$}&$_{GHz}$&$_{mJy}$\\
\hline
       J0736+4744$_e$ &q20.40r&        &  01\fs11 23\farcs1 &42.9&  53.0&   61.8&   59.2 &   48.5&   47.2&   30.0& 31.5&3.0&64\\
       J0754+3033$_e$ &q17.38r&0.796  &  48\fs72 55\farcs4 &47.8&  54.1&  170&  182 &  231&  225&  201&    142&9.3 &233 \\
       J0804+5431$_i$ &g18.33r&(0.29)  &  59\fs25 57\farcs9 &38.9&  43.4& 83.2& 81.0 & 67.5&   67.2&   49.3&38.4&5.4 &84 \\
       J0819+3823$_b$ &g21.72r&        &  00\fs81 59\farcs7 &15.8&  20.6&  104&  108&   93.2& 90.5&44.6&24.2&5.9&115\\
       J0821+3107$_e$ &q17.04r&2.624   &  07\fs61 52\farcs0 &110& 123&134&131 &  112&  109&   70.5& 48r&3.4 &140 \\
       J0905+3742$_h$ &g22.41r&        &  27\fs45 53\farcs2 &53.6&74.1&104&  101&   71.3&   67.8&   39.0& 24&3.3&113\\
       J0943+5113$_h$ &      &         &  51\fs79 21\farcs7 &73.6&  91.6&  161&  148& 69.6&   64.1&   25.8&19&3.7&209\\
       J0951+3451$_b$ &g19.42r&(0.30) &  11\fs39 31\farcs5 &19.2&  27.8&   61.1&   62.0 &56.7&   55.2&38.1& 27.6&5.1&62\\
       J0955+3335$_f$ &q17.37r&2.477  &  37\fs97 04\farcs5 &38.6&  49.4&  101&  106 &  107&  105& 70.2& 38.4&6.6&109 \\
       J1002+5701$_h$ &g22.42r&        &  41\fs65 11\farcs6 &19.8&  32.6&  135&  133&   77.2& 70.9&   20.1& 12.9&4.4&139\\
       J1004+4328$_f$ &g21.87r&(0.44)  &  31\fs32 33\farcs2 &12.3&  16.9&   38.0&38.1 & 31.9&29.8&32.8& 27.7&8.1 &36 \\
       J1008+2533$_c$ &      &         &  20\fs73 39\farcs5 &39.0&  57.6&  111&110&98.2& 96.2&   74.4&48.1&5.9 &109 \\
       J1020+2910$_c$ &      &         &  08\fs65 29\farcs7 &19.3&21.7& 24.6&23.6 &19.1&19.2&11.4&7.9&3.3 &26 \\
       J1020+4320$_f$ &q18.94r&1.962  &  27\fs18 56\farcs4 &116&163&  239&  227 &  183&  175&  135&    84.5&1.4&245\\
       J1025+2541$_c$ &      &         &  23\fs79 57\farcs7 &21.7&  23.1&   50.9&49.3 &34.2&32.3&12.6&5.7&4.1 &51 \\
       J1035+4230$_h$ &q19.21r&2.441  &  32\fs59 20\farcs0 &23.1&  29.2&  104&  109&  110&  108&   74.7&54&5.8&119\\
       J1037+3646$_f$ &g22.57r&($>$1)  &  57\fs29 55\farcs7 &66.8&  96.3&  147&  140 &   95.9&   90.5&56.3&32.5&1.6&157\\
       J1044+2959$_h$ &q18.99 &2.983  &  06\fs32 02\farcs7 &38.3&  53.8&  176&  182&  169&  167&  125&92&4.0&183\\
       J1046+2600$_c$ &      &         &  57\fs27 46\farcs2 &12.5&  15.4&   36.0&   35.0 &   27.2&25.9&13.0&5.6&  6.2&35\\
       J1047+3945$_h$ &q20.11r&($>$1)  &  03\fs24 45\farcs7 &37.8&  47.5&   55.9&   54.4 &   42.7&   42.1&29.8&24 &2.1&62\\
       J1052+3355$_f$ &q16.99 &1.407  &  50\fs09 04\farcs1 &10.4&  15.5&   46.3&   44.9 &29.9&27.6&18.8&25.8&3.0&50\\
       J1053+4610$_i$ &g22.40r&(0.51)  &  53\fs47 58\farcs9 &10.4&  15.1&   36.0&38.1 &35.1&34.8&38.1&36.5&11 &40 \\
       J1054+5058$_i$ &      &         &  13\fs96 16\farcs6 &11.8&  12.3&   22.8&27.0 &30.6&31.1&39.4&38.8&20 &39 \\
       J1058+3353$_f$ &g18.72r&0.265  &  08\fs84 04\farcs0 &21.7&  30.4&   66.9&   65.2 &52.9&51.3&44.7&  36.6&3.0&77\\
       J1107+3421$_f$ &g21.42r&(0.10)  &  34\fs31 18\farcs2 &24.6&  38.3&   77.4&   76.1 &53.9&50.7&29.3&28.5&2.7&101\\
       J1109+3831$_i$ &g21.09r&(0.12)  &  39\fs18 21\farcs5 &9.6& 17.6&   54.3&   58.5 &77.0&77.1&53.3&37.7&8.5&73\\
      J1135+3624$_h$ &s23.50r&        &  52\fs29 22\farcs5 &29.3&  38.5&   57.6&57.5 &43.0&40.8&22.4&15&  4.0 &60 \\
       J1137+3441$_h$ &q23.35r&0.835  &  09\fs09 56\farcs2 &23.6&  28.2&   70.1&71.4 &76.6&75.6&79.1& 91&14.5 &87 \\
      J1203+4803$_i$ &q16.21r&0.817 &  29\fs92 13\farcs6 &156& 191&  436&  451 &  596&  617&862& 909&34 &955 \\
       J1218+2828$_c$ &      &         &  06\fs32 25\farcs0 &18.3&  24.3&   59.9&   79.0 &94.7&94.1&60.8& 32.3&9.5&95\\
      J1239+3705$_a$ &      &         &  36\fs38 08\farcs1 &8.7?&   8.9&   86.3&95.2 &115&115&89.4&82.9&  5.8&123\\
       J1240+2425$_c$ &q16.6  &0.829   &  09\fs06 29\farcs7 &36.4&  43.4&   66.8&   62.9 &   38.7&37.1&25.3&11.5& 2.7&67\\
       J1240+2323$_c$ &     &          &  17\fs76 49\farcs9 &14.5&  19.9&54.6&56.2 &60.6&60.1&53.0& 35.1&7.9 &62 \\
       J1241+3844$_a$ &     &          &  43\fs00 03\farcs7 &19.0&  22.1&24.2&23.2 &20.3&20.9&  17.0&21.5&4.1 &23 \\
       J1251+4317$_f$ &q18.81r&1.440  &  46\fs18 27\farcs5 &13.5&  24.0&   52.3&   54.8 &50.0&49.2&43.7& 36.0&1.8&53\\
       J1258+2820$_g$ &      &         &  02\fs02 02\farcs5 &30.4&34.0&39.9& 40.5 &39.6&38.4&32.4&24.1&  4.7 &41 \\
       J1300+4352$_f$ &      &         &  20\fs18 24\farcs8 &144& 174&  235&  234 &  220&  219&205&153&5.9 &234 \\
       J1309+4047$_f$ &q19.01r&2.910  &  41\fs49 54\farcs9 &34.7&  53.4&  130&  128 &  102&   95.2&59.0&35.9& 2.0&129\\
       J1319+4851$_g$ &q19.18r&1.170  &  30\fs31 03\farcs9 &20.0&  29.2&   48.2&   45.9 &   38.2&36.8&26.7&23.2 &2.4&53\\
      J1321+4406$_g$ &g21.45r&(0.32)  &  25\fs47 33\farcs7 &33.1&  39.9&72.3&74.8 &74.1&72.6&64.7&   54.8&7.6 &76 \\
       J1322+3912$_g$ &q17.61r&2.985  &  55\fs65 08\farcs0 &108& 123&  212&  215 &  192&  189&  128&   90.0&5.8&216\\
       J1330+5202$_g$ &g21.01r&(0.57)  &  42\fs59 15\farcs3 &83.3&  97.2&  151&  154 &153&152&133&  110&6.8 &156 \\
       J1336+4735$_g$ &s19.91r&        &  11\fs69 20\farcs5 &28.8&  36.8&   60.0&   58.9 &   47.3&45.8&31.2&18.6 &1.8&60\\
       J1352+3603$_d$ &g18.15r&(0.33)  &  01\fs00 51\farcs5 &58.1&  65.4&  115&  117 &  111&  109&   72.1&44.9&7.9&118\\
       J1420+2704$_g$ &s20.37r&        &  51\fs25 25\farcs2 &8.0&  14.3&   55.5&   55.6 &52.3&51.4&37.4&25.9&2.3&57\\
       J1421+4645$_g$ &q18.08r&1.668  &  23\fs02 47\farcs5 &105& 125&  181&  180 &  164&  161&  131&  110&1.3&180\\
       J1436+4820$_g$ &g21.39r&($>$1)  &  18\fs89 39\farcs2 &17.0&  26.7&   74.1&   75.2 & 65.2&62.5&40.5&22.8&2.1&75\\
       J1459+3337$_d$ &q17.71r&0.645  &  58\fs49 00\farcs8 &13.2&  19.1&  162  &  189&374&394&612&569&17 &611 \\
       J1512+2219$_g$ &g21.08r&(0.40)  &  28\fs23 38\farcs7 &21.7&  30.4&   34.1&   31.7 &16.4&15.7&14.3&    8.1&2.3&47\\
       J1528+3816$_d$ &g20.95r&($>$1)  &  37\fs02 06\farcs4 &20.1&  23.5&47.5&50.3 &66.0&65.1&76.6&   67.2&14 &75 \\
       J1530+2705$_d$ &g14.31r&0.032  &  16\fs22 51\farcs2 &10.8&  11.1&   39.7&   44.3 &67.3&68.7&67.1&41.1 &12&76\\
       J1530+5137$_g$ &      &         &  19\fs75 30\farcs3 &42.6&  53.8&   68.3&   68.5 &   70.1&71.9&76.1&   76.6&-&-\\
       J1547+3518$_d$ &s21.37r&(0.96) &  54\fs13 42\farcs5 &10.4&  12.6&43.7&48.7 &65.8&66.2&83.9&   78.8&17 &82 \\
       J1602+2646$_d$ &g18.58r&0.372  &  39\fs63 05\farcs4 &33.3&  39.8&  125&  140 &226&231&256& 213&13 &269 \\
       J1613+4223$_d$ &s20.01r&(0.17)  &  04\fs80 18\farcs7 &35.5&  66.1&  215&  208   &131&122 &43.9&23.7 &2.8&226\\
\hline
\end{tabular}
\end{center}
\end{table*}
\setcounter{table}{1}
\begin{table*}[!t]
\caption{continue 
}
\begin{center}
\begin{tabular}{lccccccccccccc}
\hline
$name$&$ID$&z&ra-dec&
$S_{1.4}$&$S_{1.7}$&$S_{4.5}$&$S_{5.0}$&$S_{8.1}$&$S_{8.5}$&$S_{15}$&$S_{22}$&$\nu_m$&$S_m$\\
         &       & &$_{J2000}$&
\multicolumn{8}{c}{$_{mJy}$}&$_{GHz}$&$_{mJy}$\\
\hline
J1616+4632$_d$&B?      &       &03\fs75 25\farcs1 &65.7&73.0&99.3&101 & 129 &131 &157 &163&19?  &169 \\
J1617+3801$_d$&q19.20r &1.607 &48\fs50 40\farcs7 &46.0&21.3&27.5&73.7& 78.9&98.7&99.0  &  100& 28 &105 \\
J1624+2748$_g$&g21.61r &       &35\fs65 58\farcs0 &17.1&19.6&97.2&110 &173  &177.4&185  &149  &12 &198 \\
J1651+3417$_d$&g23.22r &(0.69)&42\fs34 00\farcs6 &$\sim$7.3&$\sim$9.4&47.8&52.3 &63.0&63.2&49.4&35.7&6.2&64\\
J1702+2643$_d$&s17.40r& (0.21)&09\fs65 14\farcs8 & 31.8&  34.9&   66.0&   67.4 &   76.6&   77.3&   86.2&  88.7& & \\
J1719+4804$_d$&q15.3  &1.084&38\fs31 12\farcs1 & 54.5&  68.6&  131  &  141 &  205 &  206  &  188  &149 & 11  & 229  \\
\hline
\end{tabular}
\end{center}
\end{table*}

We inspected the NVSS and FIRST images to make sure that the component in
the catalogue accounted for the whole flux density. The extended
objects (typically FRII and a few FRI or complex radio sources,
point-like in the 87GB, but resolved in the FIRST) were removed.

\section{The simultaneous radio spectra}
\label{spectra}
Simultaneous multifrequency observations are necessary to remove flat
spectrum variable sources from the sample. Our selection
criteria select also variable sources that
happened to be in a ``high'' activity state at the time of the 4.9 GHz
observation (e.g. Tinti et al. 2005).

Hence, we have observed at the VLA the whole ``faint'' sample during
several observing runs between November 1998 and February 2000 (Tab.1 )
measuring nearly simultaneous flux densities in L band
(with the two IFs at 1.365 and 1.665 GHz), C band (4.535 and 4.985
GHz), X band (8.085 and  8.485 GHz), U band (14.935 and 14.985 GHz)
and K band (22.435 and 22.485 GHz). 

Each source was observed typically for 40 seconds at each frequency in
a single snapshot, cycling through frequencies. 

For each observing run we spent one or two scans on the primary
flux density calibrators 3C286 or 3C147 or 3C48. Secondary calibrators were
observed for 1 minute at each frequency about every 25 minutes;
they were chosen aiming to minimize the telescope slewing time and
therefore we could not derive accurate positions for the radio sources
we observed.

The data reduction has been carried out following the standard procedures
for the VLA implemented in the NRAO AIPS software. Separate images
for each IF were obtained at L, C and X bands in order to improve the
spectral coverage of our data.

In general one iteration of phase-only
self-calibration have been performed before the final imaging. On the
final image we perfomed a Gaussian fit to measure the flux density of the radio source, and 
checked the total flux density to find evidence of resolved radio emission. 
Generally all the HFP candidates were unresolved by the present observations.

The r.m.s. noise levels in the image plane is relevant only
for measured flux densities of a few mJy, the major contribution comes
from the amplitude
calibration error.

The overall amplitude error is dominated by
the calibration error, and we estimate it is (1$\sigma$)
3\% at L,C and X bands, 5\% at U band and finally 10\% at K band.

\section{The ``Faint'' HFP Sample}

We derived the spectral indices between any pair of adjacent
frequencies.
We considered genuine HFPs the radio sources showing an
inverted spectral index steeper than 0.5 between 1.37 and
1.67 GHz {\bf or} between 1.67 and 4.54 GHz. The sources not
fullfilling this requirement are discarded
as genuine HFPs.
The final sample consists of the 61 HFPs listed in Table 2 with the
results of our observations.

The optical ID and redshift are from the NED database, when
available or from our optical ID on the Sloan Digital Sky Survey (SDSS)
when no other optical information is available.

We fitted the radio data with a purely analytic
function, with no physics behind, given it is used to determine only
``analytical'' quantities, namely the peak and the frequency at which
it occurs.
We used the function of an hyperbola:


$Log~S~=~aLog(\nu)~+~b~-~\sqrt{{c^2\over d^2}\times (Log(\nu)-e)^2+c^2}$

When the data did not sampled the region above the peak we simplified
the function omitting $aLog(\nu)$, thus resulting in an hyperbola
symmetric with respect to the vertical axis.

From this fitting curve we derived the spectral peak ($S_m$ and
$\nu_m$, last two columns of Table 2), representing the actual maximum, regardless the point where
the optical depth is unity, or any other physical measure.
Given the assumptions the statistical error of the fit does not represent the
real uncertainty on the estimate of $S_m$ and
$\nu_m$, and a conservative value is about $10\%$.

The analysis of the properties of the faint sample and a comparison with the
bright sample will be presented in a following paper. 

\acknowledgements{The VLA is operated by U.S. National Radio Astronomy Observatory which is a facility of the 
National Science Foundation, operated under cooperative agreement by Associated Universities, inc.}

\end{document}